# Monolithic Platform for Integrated Quantum Photonics with Hexagonal Boron Nitride

Milad Nonahal[1], Chi Li[1*], Haoran Ren[3], Lesley Spencer[1], Mehran Kianinia[1], Milos Toth[1,2], Igor Aharonovich[1,2*]

1. School of Mathematical and Physical Sciences, Faculty of Science, University of Technology Sydney, Ultimo, New South Wales 2007, Australia

2. ARC Centre of Excellence for Transformative Meta-Optical Systems (TMOS), University of Technology Sydney, Ultimo, New South Wales 2007, Australia

3. School of Physics and Astronomy, Monash University, Clayton, Victoria 3800, Australia

Email: chi.li@uts.edu.au; igor.aharonovich@uts.edu.au

## Abstract

*Integrated quantum photonics (IQP) provides a path to practical, scalable quantum computation, communications and information processing. Realization of an IQP platform requires integration of quantum emitters with high quality photonic circuits. However, the range of materials for monolithic platforms is limited by the simultaneous need for a high-quality single photon source, high optical performance and availability of scalable nanofabrication techniques. Here we demonstrate the fabrication of IQP components from the recently emerged quantum material hexagonal boron nitride (hBN), including tapered waveguides, microdisks, and 1D and 2D photonic crystal cavities. Resonators with quality factors greater than 4000 are achieved, and we engineer proof-of-principle complex, free-standing IQP circuitry fabricated from single crystal hBN. Our results show the potential of hBN for scalable integrated quantum technologies.*

Integrated quantum photonics (IQP) offers a promising pathway to large scale quantum information technologies.[1-3] The main constituents of quantum circuitry are quantum emitters that ideally exhibit a spin-photon interface, and photonic elements that include waveguides, cavities and resonators[4, 5]. The overarching goal of the IQP approach is based on coupling optically active emitters to nanophotonic constituents and achieve coherent interactions between the quantum elements and photonic cavities[6-11]. Over the last decade, a significant effort has been devoted to key quantum experiments, including Purcell enhancement, single photon switching and non-linear effects at the single photon level, using a variety of material systems.[12-22] However, the ultimate solid-state platform for on-chip IQP circuitry is yet to be established.

Hexagonal Boron Nitride (hBN) has recently emerged as a promising platform for integrated quantum photonics[23]. hBN hosts numerous bright quantum emitters that can be engineered on demand, as well as a range of optically active spin defects with clear optically detected magnetic resonance signatures. Critically, hBN is a wide bandgap van der Waals material that can easily be exfoliated from its bulk form to a range of thicknesses from a monolayer to a few hundred nanometers to support light propagation across a broad spectral range. This unique combination of photophysical properties spurred interest in engineering quantum photonic circuitry from hBN. To date, proof of principle demonstrations of isolated photonic resonators and basic optical elements fabricated from hBN have been reported.[24-31] However, a holistic approach to chip-scale nanofabrication is yet to be unveiled, predominantly, due to challenges in fabrication of hBN as it is a robust material that is chemically inert.

Here we demonstrate the feasibility of fabricating on-chip and free-standing IQP elements from hBN. We focus on the monolithic approach, where entire devices are fabricated entirely from hBN. This approach has numerous advantages, including the ability to minimize coupling losses and to position quantum emitters at the highest field of a cavity.[32] We start by addressing the key issue of hBN fabrication. Next, we engineer all the necessary components of a potential quantum circuit - including waveguides, couplers, splitters and photonic crystal cavities (PCCs). Finally, to illustrate the full potential of this platform, we demonstrate a fully integrated monolithic IQP chip made entirely of suspended hBN.

The IQP fabrication process is illustrated schematically in Figure 1a. Mechanically exfoliated hBN flakes of desired thickness are first identified using an optical microscope and their thicknesses are verified by atomic force microscopy (AFM). A selected flake is then spin-coated with a thin layer of electron-sensitive resist and patterned using electron beam lithography (EBL). After development, the pattern is transferred onto the flake via induced coupled plasma reactive ion etching (ICP/RIE) in a reactive $SF_6$ gas environment. The sample was immersed in KOH solution to selectively undercut the supporting substrate straight after removing the residual polymer. Figure 1b shows a schematic illustration of a monolithic hBN IQP chip composed of waveguides (photon input), PCCs and resonators (qubit processing), and out-couplers for photon detection. In this work, we demonstrate the ability to fabricate from hBN all basic components of quantum circuitry, as is illustrated explicitly in Figure 1c (i - iii). These include waveguides, beamsplitters, resonators, optical cavities and out-couplers used to direct photons to detectors.

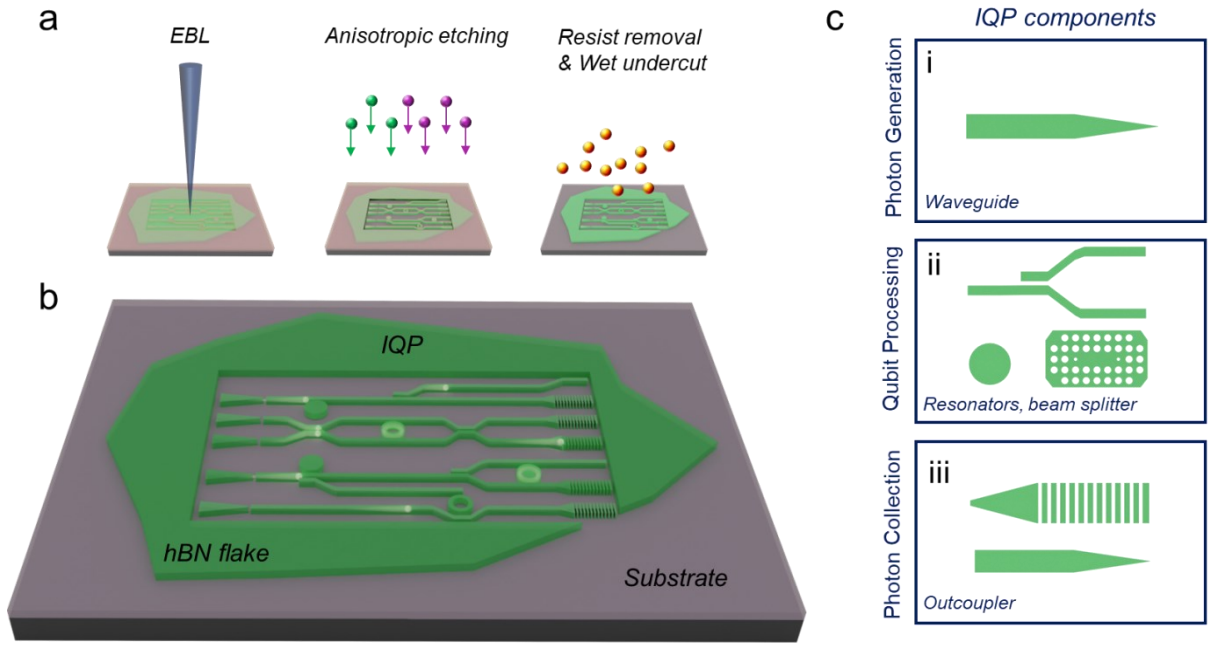

*Figure 1. Schematic illustration of monolithic fabrication using hBN.* (a) Key steps of the fabrication process: electron beam lithography, reactive ion etching and resist removal. (b) Integrated circuit fabricated from single crystal hBN. (c) IQP components: (i) low-loss waveguide (ii) resonators, optical cavities and beam splitter (iii) out-couplers for detection. Two out-couplers are shown, for top and side collection of photons.

Details of the hBN fabrication process are provided in the methods section. Here we emphasize the key elements needed to achieve high quality optical components - namely, anisotropic etching and the use of a hard mask that achieves a smooth and vertical sidewall morphology. Table 1 summarizes the etch parameters used in the current work, and those used in prior works on fabrication of hBN nanophotonic structures. As is demonstrated below, our process yields consistent, smooth etch profiles in a variety of nanostructures. This is achieved primarily by employing a relatively low reactive gas partial pressure and by maximizing ion directionality. To achieve this, a low $SF_6$ concentration of 1.5 % was employed, the chamber pressure was minimized, and ICP power was also minimized. Under these conditions, ion collisions, ion reflection and reactivity of the plasma are all suppressed, thereby inhibiting lateral etching of hBN. Lateral etching produces sloped sidewalls and can give rise to surface roughening, both of which are detrimental to photonic structures and must be minimized.

Our RIE parameters result in very slow etching of the e-beam resist. This makes the resist an excellent hard mask for single-step patterning hBN, thus eliminating the need for a separate liftoff step which can give rise to sidewall roughening. In addition, metal erosion in a highly reactive $SF_6$ environment can also cause roughening, which is avoided in our process. We note that our etch rate of 5 nm/s is suitable for chip-scale fabrication, and on par with those of other material systems.

|  | mask | Chamber Pressure | ICP/RIE power | SF6 concentration | Etch rate |
|---|---|---|---|---|---|
| This work | resist | 1 mT | 1/300 W | 1.5% | 5 nm/s |
| Froch et. al[28] | resist | 1 mT | 50/100 W | 10% | - |
| Kuhner et. al[33] | metal | 6 mT | 150/300 W | - | 4 nm/s |
| Kim et. al[24] | resist | 10 mT | 0/100 W | 100% | ~ 13 nm/s |

*Table 1. Anisotropic dry etch conditions used here and RIE parameters used in prior works on etching of hBN.*

Employing the conditions described above, we fabricated a number of optical components from hBN. First, we engineered waveguides. Optical images of a hBN flake before (after) the etching process are shown in Figure 2 a(b). Magnified scanning electron microscope images of the taper head and tail sections of the waveguide are shown in Figure 2c. A significant advantage of working with a layered material such as hBN is the ability to transfer the structures between fabrication steps. To this end, the chosen hBN flake with the correct thickness was first identified and patterned in the middle of the substrate, where the fabrication step (resist spin coating) does not generate edge artifacts. After developing the pattern, the hBN and the resist mask were transferred to the edge of the substrate, so that the tapered waveguides can be effectively used. Finally, the pattern was transferred to the hBN using ICP/IRE.

Functionality of the waveguides is demonstrated by optical measurements. The waveguide was excited by a lensed fiber through the tapered area and the light was collected from the tail end of the waveguide as is seen in the bright-field image shown in Figure 2d. Due to the high quality of the fabrication process, minimum scattering is observed in the body of the waveguide, resulting in efficient light confinement and guiding.

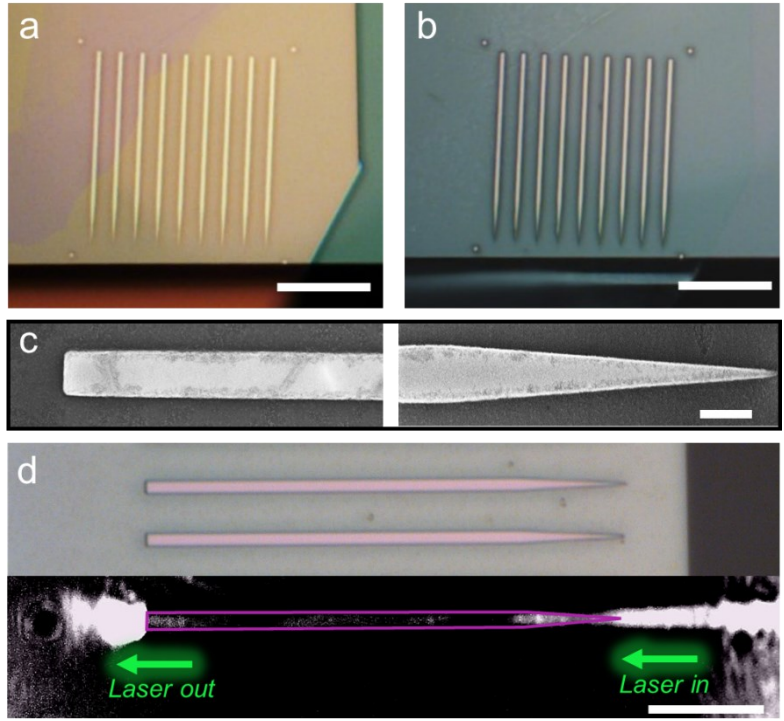

*Figure 2. Fabrication and characterization of optical waveguides.* Optical images of the (a) pre- and (b) post-fabricated array of tapered waveguides showing that the structures are fully etched after the RIE step. The scale bar corresponds to 25 µm. (c) SEM image of two sections of the waveguides. The scale bar is 1 µm. (d) magnified optical microscope image of the tapered waveguides illuminated by white light (top) and laser (bottom). The scale bar is 10 µm.

Now we turn to photonic resonators. We focus on three device geometries that are most common in integrated nanophotonics. These are microdisk cavities, and 1D and 2D photonic crystal cavities. Note that suspended cavities are preferred in order to maximize refractive index contrast and hence light confinement within the cavities. This is typically achieved by undercutting a sacrificial layer on which the cavity material is grown epitaxially. In our case, we transfer exfoliated hBN flakes onto silicon, and undercut the silicon using a KOH wet chemical etch process.

Figure 3a shows a tilted, false-color SEM image of a suspended disk resonator (3 µm in diameter) which is supported by a small silicon pillar indicating the feasibility of wet chemical etching to fabricate such small free-standing structures. 10 wt% KOH was used for the Si substrate undercut (refer to method). For optical characterization, we employed a standard confocal microscope with a 532 nm excitation source. A PL spectrum from the hBN microdisk is shown in Figure 3b. It exhibits a clear set of whispering gallery mode (WGM) resonances, seen as sharp peaks in the spectrum. To obtain the quality factor $(Q = \lambda/\Delta\lambda)$, a high-resolution PL spectrum of an individual WGM was recorded and the peak was fitted by a Lorentzian function. The microdisk exhibits a narrow linewidth of 180 pm and a high Q of >3000 (Figure 3c). This represents a 6-fold improvement in *Q*-factor over prior reports.[25] The improvement is predominantly due to the experimental realization of a suspended microdisk resonator from hBN, as well as the improved etch recipe.

Figure 3d is a top-view, false-color SEM image of a 1D hBN PCC. The cavity consists of an array of holes in a waveguide, with a varying diameter (down to 80 nm). The cavity resonance is shown in Figure 3(e, f). The resonance width is 140 pm, which corresponds to a Q factor of >4000 (Figure 3f). Next, we fabricated 2D PCCs. Figure 3g shows an SEM image of a 2D PCC structure (L5 configuration, whereby 5 holes are linearly eliminated from the central area). A close-up of the cavity center is shown in the inset. The 2D PCC exhibits a single optical mode (Figure 3h) with a modest Q factor of ~ 2000, as is shown in the high-resolution spectrum in Figure 3i. The modest quality factor for the 2D configuration is caused by losses within hBN due to its relatively low refractive index. Nevertheless, for some applications, 2D PCC geometries can be advantageous for scaling up to an on-chip IQP circuitry.

Purcell enhancement, given by $F_p = \frac{3}{4\pi^2}(\frac{\lambda}{n})^3 \frac{Q}{V}$, for the above PCCs, can potentially be very high. Given the low mode volume of the 1D PCCs *(V~ 1.5(λ/n)³)*, and the *Q factor of ~ 4000,* the Purcell enhancement is expected to exceed 200. Note that for monolithic cavities fabricated entirely from hBN, the emitter would be positioned within the high field region of the cavity, rather than coupled evanescently. This should result in substantial overlap with the cavity field, and hence a significant enhancement.[6, 34, 35]

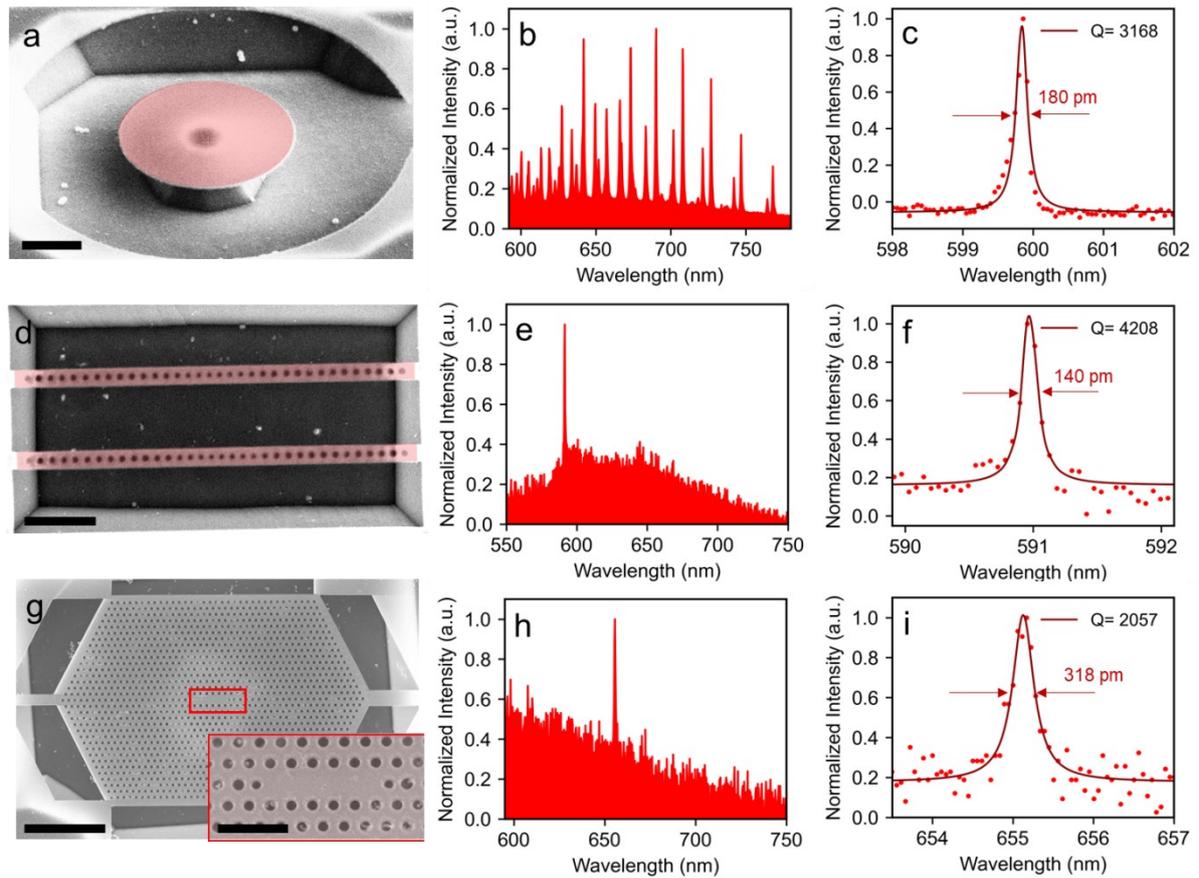

*Figure 3. Fabrication and characterization of photonic resonators. (a, d, g) False-color SEM images of a suspended hBN microdisk resonator, 1D PCC and 2D PCC, respectively. The scale bars in (a, d) and (g) correspond to 1, 1, and 4 µm, respectively. (g) Inset - magnified SEM image of the 2D PCC center area. The scale bar corresponds to 1 µm. (b, e, h) PL spectra showing WGMs in the microdisk cavity, and resonances in the 1D and 2D PCCs. (c, f, i) High resolution PL spectra showing the widths of the optical cavity modes. The data were fitted with Lorentzian functions to obtain the resonance quality factors ($\lambda/\Delta\lambda$).*

Finally, we fabricated a proof-of-concept suspended IQP circuit from a single crystal hBN flake. Figure 4a, shows an optical image of the hBN IQP circuit. The exfoliated hBN flake on a silicon substrate is sufficiently large to accommodate the 50 µm x 25 µm circuit, seen as a rectangular region in the optical image. The contrast difference is caused by the selective undercut etch process - the region underneath the circuit was chemically etched by KOH solution, admitted to the silicon substrate through the large rectangular holes seen in Figure 4d. The platform consists of air hole arrays in a hBN slab, specifically a hexagonal lattice with a pitch distance of 300 nm and air holes with a radius of 100 nm. One row of holes was eliminated to perform the waveguide function. For the tapered waveguide, we used triangle couplers with air to taper down the cross section of the guided wave inside the photonic crystal waveguide.

The electromagnetic field guiding in the photonic crystal waveguides were numerically simulated in the COMSOL Multiphysics software. The electric field distributions of the guided waves are given in Figure 4b,c, wherein a linear polarized electric field input (out-of-plane) was placed at the left side of the waveguides. Low-loss wave guiding and propagation in both tapered and straight waveguide structures can be achieved based on our simulation results. Figure 4d shows a top-view SEM image of the IQP circuit. Three main elements are shown on the image, namely data generation (red box), manipulation (cyan) and detection (green). Figure 4(e – g) shows the higher magnification SEM images of these sections. Note that the IQP chip was not designed for any specific operation. Rather it shows that all the required components - including couplers, waveguides, PCCs, routers, beam splitters and out-couplers can be engineered in a single, suspended flake of hBN using a robust nanofabrication method.

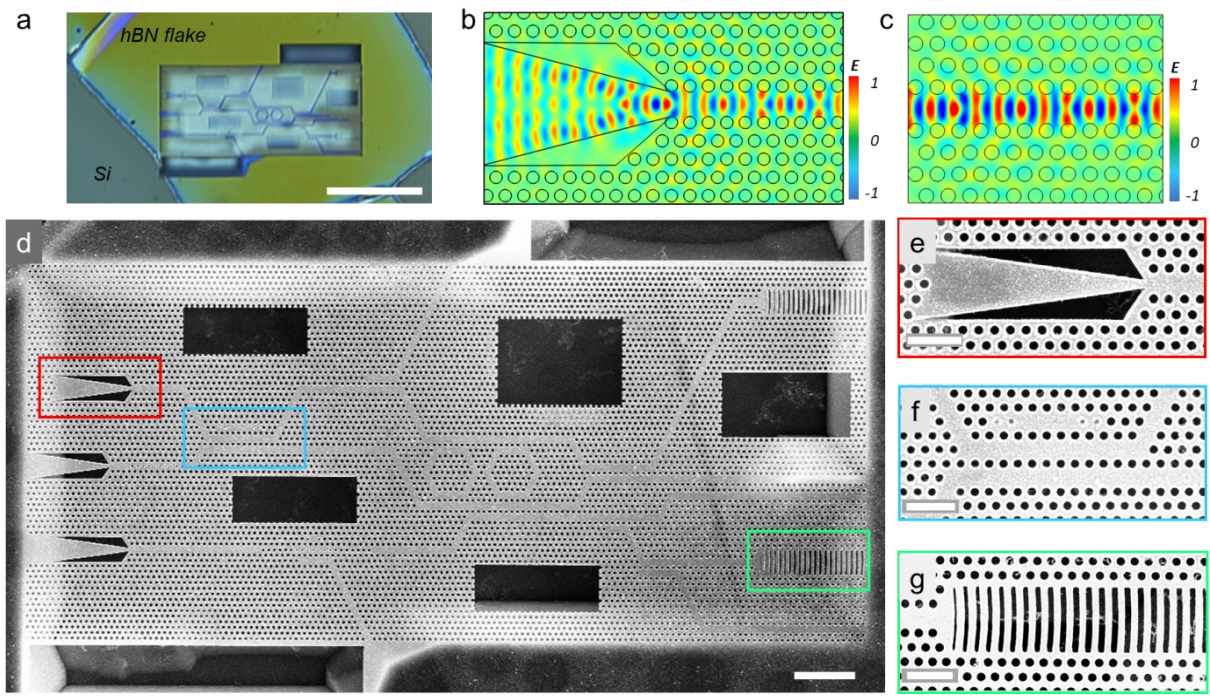

*Figure 4. Suspended on-chip IQP circuit fabricated from single crystal hBN. (a) Optical image of the suspended structure embedded in an exfoliated hBN flake. The scale bar corresponds to 20 µm. Simulated electric field distribution of tapered (b) and straight (c) photonic crystal waveguides. For both simulations, an out-of-plane electric field is incident on the left side of the waveguide structure at a wavelength of 600 nm. (d) top-view SEM image of the circuit. The scale bar corresponds to 3 µm. Three elements are highlighted by Red, Cyan and green boxes and zoomed-in images shown in (e, f, g), corresponding to photon generation, manipulation and detection, respectively. The scale bars in (e-g) corresponds to 1 µm.*

To summarize, we developed a reliable and robust recipe to engineer quantum photonic circuitry from hBN. Specifically, we realized microdisk cavities, 1D and 2D PCCs, and an entire suspended monolithic IQP chip fabricated from hBN. The suspended architecture yields improved light confinement, with measured *Q* factors in excess of 4000 for 1D PCC. Our work marks an important step towards integration of quantum emitters and scalable quantum photonic elements in hBN. It demonstrates a promising pathway to realization of advanced spin - photon interfaces, and deployment of hBN in quantum optomechanics, and in applications based on strong coupling between a quantum emitter and a photonic resonator. Using an on-chip set of low-loss waveguides and optical cavities, IQP can provide coherent quantum circuitry capable of performing core quantum tasks including quantum state generation, manipulation and detection of information.

**Acknowledgement**

This work is supported by the Australian Research Council (CE200100010, DE220101085, DP220102152) and the Office of Naval Research Global (N62909-22-1-2028).